\documentclass[12pt]{article}
\usepackage{graphics}
\newcommand{\bq}{\begin{equation}}
\newcommand{\eq}{\end{equation}}
\newcommand{\bqa}{\begin{eqnarray}}
\newcommand{\eqa}{\end{eqnarray}}
\newcommand{\B}{\beta}
\newcommand{\G}{\gamma}
\newcommand{\Bbb}{B^{}_{\B\B}}
\newcommand{\Bbg}{B^{}_{\B\G}}
\newcommand{\Bgg}{B^{}_{\G\G}}
\newcommand{\Sg}{\sin\!3\G}
\newcommand{\dbe}{\partial_{\B}}
\newcommand{\dga}{\partial_{\G}}
\newcommand{\rnw}{\sqrt{r\over w}}
\newcommand{\nn}{\nonumber}
\newcommand{\mod}{{\, \rm mod \, }}
\newcommand{\delIe}{\delta_{I\, {\rm even}}}

\begin{document}

\title{\bf Collective Quadrupole Excitations\\
in the $50< Z,\, N< 82$ Nuclei\\
with the Generalized Bohr Hamiltonian}

\author{L. Pr\'ochniak, K. Zaj\c ac and K. Pomorski\\
{\it Institute of Physics, The Maria Curie-Sk{\l}odowska University,}\\
{\it pl. M. Curie-Sk{\l}odowskiej 1, 20-031 Lublin, Poland}\vspace{0.5cm}\\  
S. G. Rohozi\'nski and J. Srebrny\\
{\it Department of Physics, Warsaw University,}\\
{\it Ho\.za 69, 00-681 Warsaw, Poland}}

\date{}
\maketitle

\begin{abstract}
The generalized Bohr Hamiltonian is applied to a description of low-lying
collective excitations in even-even isotopes of Te, Xe, Ba, Ce, Nd and Sm.
The collective potential and inertial functions are determined by means of
the Strutinsky method and the cranking model, respectively. A shell-dependent
parametrization of the Nilsson potential is used. An approximate 
particle-number projection is performed in treatment of pairing correlations.
The effect of coupling with the pairing vibrations is taken into
account approximately when determining the inertial functions. The
calculation does not contain any free parameter.  
\end{abstract}

\section{Introduction}
\label{intr} 
For many years the Generalized Bohr Hamiltonian (GBH)
\cite{Bo52,Ku67,EG87} has been still forming the only one model which,
having possibilities to be derived microscopically 
(e.g. \cite{Ka76,Ro77,Go85}), pretends to be able to describe
collective excitations in the transitional nuclei or nuclei soft with
respect to the quadrupole deformations. This is because the GBH is a
scalar under rotations and, thus, possesses eigenstates of a good
angular momentum whereas the most of other approaches deal with {\em
intrinsic states} of an indefinite spin. Also, a rotation--vibration
coupling, so important for transitional nuclei, is automatically taken
into account in the Hamiltonian.

The aim of the present paper is to investigate the low-lying collective
states in even--even nuclei from the region of $50< Z,\, N <82$ within
framework of the GBH. This region of the chart of nuclides is a good
laboratory to study the collective model in its most general form as
the standard approximations of it like the rotational and vibrational
models are apparently not applicable to nuclei in question. Since the
early calculation of potential energy surfaces by Arseniev {\em et al.}
\cite{Ar69} the $50< Z,\, N <82$ nuclei are referred to as gamma--soft
or susceptible to deformations leading to triaxial shapes. This is why
the Wilets--Jean model was applied already long time ago to interpret
experimental data \cite{Ro73}. On the other hand, the properties of
these nuclides have been explained within the Davydov--Filippov model
treating nuclei as rigid triaxial ones (cf. \cite{Sr93}). The
triaxiality of the Xe and Ba isotopes has been discussed once more in
the recent work of Meyer {\it et al.} \cite{Me97}. Other theoretical
approaches have also been applied to the investigated region. The
closest to our study are the applications of Kumar's dynamical
deformation model to the Te nuclei \cite{Subber} and the Frankfurt
general collective model to the Ba isotopes \cite{Petkov}. Among
methods different from ours we mention the microscopic calculations
\cite{Ha86} of number- and spin-projected two-quasiparticle states
with realistic interactions (MONSTER) and the Fermion Dynamical
Symmetry Model \cite{fdsm}, both applied to the Xe and Ba isotopes. The
cranked Hartree--Fock--Bogolyubov calculations of the yrast bands and
E2 properties have been performed for the Te and Xe isotopes
\cite{devi1,devi2}. Within the Interacting Boson Model (IBM-1) the Xe
and Ba nuclei have been treated in the SO(6) limit \cite{Ca85}. Also,
the proton-neutron Interacting Boson Model (IBM-2) has been applied to
the Xe, Ba and Ce isotopes \cite{Pu80}. We cite here only the newer
papers on broader ranges of nuclides from the discussed region.

In Section \ref{cm} we recapitulate briefly the GBH recalling all
necessary definitions and formulae. A new basis for the diagonalization
of collective Hamiltonian is presented in Section \ref{basis}. Some
details of the basis construction are given in Appendix. To construct
the collective Hamiltonian we use a standard microscopic model:
nucleons in a deformed single particle potential interacting {\em via}
the monopole pairing forces. We take the Nilsson potential with the
parametrization of Seo \cite{Se86} for a description of the
single--particle motion. In treatment of the pairing forces we utilize
an approximate projection technique based on the Generator Coordinate
Method (GCM) for projecting the BCS wave function on the correct
particle number \cite{Go86}. Then, we calculate the collective
potential by means of the Strutinsky macroscopic--microscopic method
\cite{Ni69} and the inertial functions by means of the cranking model
\cite{Ka76}. All the microscopic procedure is presented in Section 
\ref{micr}. We know for a long time that the pairing correlations affect
strongly the collective excitations. In early calculations for Xe and
Ba isotopes we have decreased artificially the strength of the pairing
interaction in order to simulate the coupling between the collective
quadrupole and pairing vibrations \cite{Ro77}. Nowadays, when ampler
data are available, we know that the effect cannot be approximated
satisfactorily in that simple way. Sakamoto \cite{Sa95} has tried to
remedy description of the collective states in xenon isotopes by
including in a static way the quadrupole pairing interactions. We do
believe that it is a dynamical effect. Here, we approximate the effect
of coupling with the pairing vibrations by calculating the most
probable dynamical value of the pairing energy gap as a function of
deformation and putting it to microscopic calculations of the
collective inertial functions. The procedure is discussed in Section
\ref{zppv}. This way we have constructed the collective Hamiltonian and
have solved it as it stands with no free parameters to be fitted. The
results for the Te, Xe, Ba, Ce, Nd and Sm isotopes are presented in
Section \ref{results}. Conclusions from our research are drawn in
Section \ref{concl}.

\section{Recapitulation of the collective model}
\label{cm}

The five dynamical variables of the model constitute a quadrupole
tensor and are usually denoted as $\alpha_\mu, \, \mu=-2,-1,\ldots ,2$.
In order to separate rotations from vibrations, the variables
$\alpha$'s are parametrized in terms of two deformation parameters,
$\B$ and $\G$ and three Euler angles, $\phi,\, \theta,\, \psi \, ({\rm
or}\; \Omega,$ in short), defining the orientation of the intrinsic
principal axes with respect to the laboratory frame:
\bq\label{bg}
\alpha_\mu = D^2_{\mu 0}(\Omega )\B\cos\G 
+\frac{\displaystyle 1}{\displaystyle \sqrt{2}}
\left(D^2_{\mu 2}(\Omega ) + D^2_{\mu -2}(\Omega )\right) \B\sin\G \ ,
\eq
where $D^\lambda_{\mu\nu}(\phi ,\theta ,\psi )$ are the Wigner functions.
Eq. (\ref{bg}) defines the intrinsic frame up to 24 rotations forming the
octahedral group O \cite{Ha62,Co76}. In consequence, only $\B$ is a 
unique function of $\alpha$'s. The remaining intrinsic variables undergo 
transformations of the group~O.

The collective Hamiltonian, when expressed in terms of the intrinsic
variables, is decomposed as
\bq\label{h}
\hat{\cal H}_{\rm coll}= \hat{\cal T}_{\rm vib}(\B ,\G) + 
\hat{\cal T}_{\rm rot}(\B ,\G ,\Omega )
 + V_{\rm coll}(\B ,\G) \ ,
\eq
where $V_{\rm coll}$ is the collective potential, the kinetic
vibrational energy reads
\bqa\label{vib}
\nn
\hat{\cal T}_{\rm vib}=-{{\hbar}^2\over{2\sqrt{wr}}}\bigg\{ {1\over \B^4}\bigg[ 
\dbe \bigg(
\B^4\rnw \Bgg\dbe\bigg) - \dbe \bigg(\B^3\rnw \Bbg\dga\bigg)\bigg]+ 
&&\\
+ {1\over \B\Sg}\bigg[ -\dga \bigg( \rnw \Sg\Bbg\dbe\bigg) +
{1\over\B}\dga \bigg(\rnw\Sg\Bbb\bigg)\dga\bigg] \bigg\} 
\eqa
and the rotational energy is
\bq\label{rot}
\hat{\cal T}_{\rm rot}={1\over 2}\sum_{k=1}^{3} \hat{I}^2_k/{\cal J}_k \ .
\eq
The mass parameters (or vibrational inertial functions) are denoted as 
$\Bbb$, $\Bbg$ and $\Bgg$. They are, in general, functions of $\B$ and 
$\G$. Moments of inertia ${\cal J}_k,\, (k=1,2,3)$, in general also
$(\B ,\G )$-dependent, are conventionally expressed by rotational
inertial functions $B_k$ : 
\bq
{\cal J}_k=4B_k(\B,\G)\B^2\sin^2(\G -2\pi k/3) \quad {\rm for~~~} k=1,2,3.
\eq
The intrinsic components of total angular momentum are denoted as $\hat{I}_k,\, 
(k=1,2,3)$. The determinants of the vibrational and rotational tensors are:
\bq\label{wr}
w=B_{\B\B}B_{\G\G}-B^2_{\B\G} ,\quad r=B_1B_2B_3 .
\eq 
The volume element has to have the form
\bq\label{meso}
{\rm d}\tau =\B^4\sqrt{wr}|\Sg| {\rm d}\B{\rm d}\G{\rm d}\Omega
=\sqrt{wr}{\rm d}\tau_0
\eq
in order to fulfill the hermiticity condition for $\hat{\cal H}_{\rm coll}$.
The collective electromagnetic E2, M1 and E0 operators are determined by their
$(\B ,\G )$-dependent intrinsic components in the following way:
\bq\label{mult}
\begin{array}{rl}
\hat{\cal M}({\rm E2};\mu)&=\sqrt{5\over 16\pi}\big\{ D^2_{\mu 0}(\Omega)
Q_0(\B,\G)\\
&+ {1\over \sqrt{2}}\left(D^2_{\mu 2}(\Omega ) + D^2_{\mu -2}(\Omega )\right)
Q_2(\B,\G)\big\},\\ 
\hat{\cal M}({\rm M1};\mu)&=\sqrt{3\over 4\pi}\big\{D^1_{\mu 0}(\Omega)
{\cal G}_3(\B,\G)\hat{I}_3\\
&- {1\over \sqrt{2}}\left(D^1_{\mu 1}(\Omega ) - D^1_{\mu -1}(\Omega )\right)
{\cal G}_1(\B,\G)\hat{I}_1\\
&+{i\over \sqrt{2}}\left(D^1_{\mu 1}(\Omega ) + D^1_{\mu -1}(\Omega )\right)
{\cal G}_2(\B,\G)\hat{I}_2\big\},\\
\hat{\cal M}({\rm E0})&= ZeR^2(\B,\G).
\end{array}
\eq

Although the above description of the nuclear collective excitations is
sometimes called {\em a geometrical model} it is needless at this stage 
to refer to the geometrical picture of nucleus. The model is just defined
by seven functions of $\B$ and $\G$: $V_{\rm coll}$, the collective potential
and $B_{\B,\B},\, B_{\B,\G},\, B_{\G,\G},\, B_1,\, B_2,\, B_3,$ the inertial
functions. To investigate electromagnetic properties additional functions
should be defined, for instance, $Q_0$ and $Q_2$, the quadrupole electric
moments, ${\cal G}_1$, ${\cal G}_2$ and ${\cal G}_3$, the gyromagnetic
functions and $R^2$, the square of electric radius, as presented in eq.
(\ref{mult}) above. All of these 13 functions have to be determined
from a microscopic theory. The geometry may conceivably be inherent in
there.

\section{The basis in the collective space}
\label{basis}

The methods used hitherto to solve the eigenvalue problem for
Hamiltonian of eq.~(\ref{h}) can be divided into two groups. The former
group consists in a direct numerical solution of a system of partial
differential equations using finite difference methods
\cite{Ku67,Ro77,Tr92}. The latter methods resolve themselves to matrix
eigenvalue problems by means of expanding the eigenfunctions in a
complete set of functions of $\B ,\, \G ,\, \phi ,\, \theta ,\, \psi$
\cite{Du70,Gn71,Ku74,Li82}. Then, the problem of construction of an
appropriate basis for a given angular momentum arises. The five-dimensional
oscillator wavefunctions of a good angular momentum constitute a
natural basis \cite{Co76,Ch76,Ch77,Gh78,Sz80}. 

Following the idea of Libert and Quentin we construct the basis 
\cite{Li82} from a complete set of square integrable functions of
$\B_0= \B \cos{\G},\, \B_2= \B\sin{\G}$ and $\Omega$ in the form of product
of gaussians and monomials in $\B$'s, and the Wigner function in $\Omega$:
\bq
e^{-\mu\B^2/2}(\B_0)^{n_0}(\B_2)^{n_2}D^{I}_{ML}(\Omega)
\eq
for $n_0,n_2 = 0,1,\ldots ,\, I = 0,1,\ldots$ and $M,L = -I,\ldots ,I$.
This basis is equivalent to the set of functions 
\bq\label{set1}
{\varphi}^{IM}_{L m n}(\B ,\G ,\Omega )=
e^{-\mu \B^2/2}\beta^n\bigg\{\begin{array}{c}
\cos m\G\\
\sin m\G
\end{array}\bigg\}
D^{I}_{ML}(\Omega)
\eq
with $n=0,1,\ldots$ and $ m=n,n{-}2,\ldots, 0 \mbox{ or }1$.
The functions ${\varphi}^{IM}_{L m n}(\B ,\G ,\Omega )$ form a complete
 set of nonorthogonal, square integrable functions with respect to measure 
${\rm d}\tau_0$. The basis functions (\ref{set1}) depend on a parameter
$\mu$ which is to be chosen further. Then we symmetrize
appropriately the functions $\varphi$ of eq. (\ref{set1}) under the octahedral
group O using a method close to that presented in ref. \cite{De89}.
Details of the symmetrization procedure are published elsewhere \cite{Pr98}.
The O-invariant functions generated from the set of eq. (\ref{set1}) and 
forming a basis take the following form:
\bq\label{set4}
\tilde{\Psi}^{IM}_{Lmn}(\B,\G,\Omega)=
e^{-\mu\B^2/2}\B^{n}\sum_{K={\rm even}\ge 0}
\tilde{f}^{I}_{LmK}(\G)\Phi^{I}_{MK}(\Omega)
\eq
with ranges of $I,\, M,\, n$ and $m$ given above. Quantum number $L$
runs over natural even values within a range specific for given $I$
and $m$ \cite{Pr98}. The form of functions $\tilde{f}^I_{LmK}(\G )$
and possible values of $L$ are given in Appendix. The Euler angles
dependence of $\tilde{\Psi}$'s is given through the Wigner functions:
\bq\label{symD}
\Phi^I_{MK}(\Omega) = \sqrt{2I+1\over 16\pi^2(1+\delta_{K0})}
\left(D^I_{MK}(\Omega )+ (-1)^ID^I_{M-K}(\Omega)\right)
\eq
The nonorthogonal basis of eq. (\ref{set4}) can be used to diagonalize
Hamiltonian $\hat{\cal H}_{\rm coll}$ as, for instance, Kumar does
\cite{Ku74}. Instead, we orthogonalize it applying
Cholesky--Banachiewicz procedure \cite{Ra65} as is described in
Appendix. In order to make the basis most effective the parameter $\mu$
has to be chosen appropriately. The natural choice is to fit potential
$V_{\rm coll}$ to a paraboloid of revolution with stiffness $C$, to
take a mean value $B$ of all of inertial functions and then to use the
oscillator formula: 
\bq
\mu = \left(\frac{\displaystyle BC}{\displaystyle \hbar^2}\right)^{1/4}.
\eq
We have tested the above choice of $\mu$ and checked that it works
fairly well. The maximal value of the order of the polynomial $\B^n$ in
eq. (\ref{set4}) is fixed by the plateau condition. In our case it was
$n=16$. 

\section{The microscopic model}
\label{micr}

We describe microscopically the nucleus as a system of nucleons moving
in a deformed mean field and interacting through monopole
(state-independent) pairing forces. Thus, the microscopic Hamiltonian
reads: 
\bq\label{mh}
\hat{H} = \hat{H}_{\rm s.p.} + \hat{H}_{\rm pair}
\eq
with
\bqa
\label{sp}
&& \hat{H}_{\rm s.p.}= \sum_{\nu >0}\langle\nu|\hat{h}|\nu\rangle 
(c^+_{\nu}c_{\nu} + c^+_{-\nu}c_{-\nu})\ ,\\
\label{pair}
&& \hat{H}_{\rm pair} = -\sum_{t=\pm 1/2}G_t\sum_{\nu >0}
\langle\nu|(\frac{1}{2} + t\tau_3)|\nu\rangle c^+_{\nu}c^+_{-\nu}
\sum_{\nu'>0}
\langle\nu'|(\frac{1}{2} + t\tau_3)|\nu'\rangle c_{-\nu'}c_{\nu'},
\eqa
where $t=\pm 1/2$ is the isospin projection of neutron and proton, 
respectively. We use the Nilsson single particle Hamiltonian $\hat{h}$ 
with the shell dependent parametrization \cite{Se86} of the 
{\bf ls} and {\bf l}$^2$ correction terms. Namely
\bqa
&&v_{\bf ls}=-2\hbar\omega_{0}\kappa_{Nl}{\bf l}\!\cdot\!{\bf s},\\
&&v_{{\bf l}^2}=-\hbar\omega_{0}(\mu_{Nl}{\bf l}^2- {<}\mu_{Nl}{\bf 
l}^2{>}_{\rm shell}),
\eqa
where
\bqa
&&\kappa_{Nl}=\kappa_{0}[1+8\mu_{Nl}(N+3/2)]+\kappa_{1} A^{1/3}P_{s,Nl},\\
&&\mu_{Nl}=\mu_{0}P^2_{i,Nl}
\eqa
and $P_{s,Nl}$, $P_{i,Nl}$ are the following integrals (evaluated
numerically) of the harmonic oscillator radial wave function ${R}_{Nl}(r)$:
\bqa
&&P_{i,Nl}=\int_{0}^{R_{0}-a/2}{R}^2_{Nl}(r)r^2 dr\\
&&P_{s,Nl}=\int_{R_{0}-a/2}^{R_{0}+a/2}{R}^2_{Nl}(r)r^2 dr
\eqa 
Here $R_0$ and $a$ stand for the nuclear radius and the surface thickness,
respectively. We take, after ref. \cite{Se86}, $\kappa_{0}=0.021$,
$\kappa_{1}=0.9$, $\mu_{0}=0.062$ and we take $A=130$ as the average
nuclear mass in the considered region.

\begin{figure}[htb]
\begin{center}
\scalebox{0.5}{\includegraphics{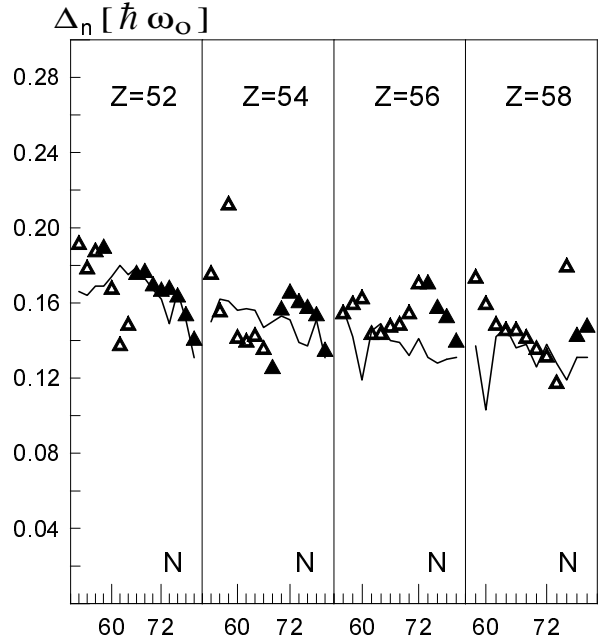}}
\caption{\footnotesize The theoretical and experimental neutron energy
gaps for isotopes of Te, Xe, Ba and Ce. Experimental points obtained
from measured or estimated \cite{Mo93} nuclear masses are represented
by full or open triangles, respectively. Theoretical values are marked
with lines.}\label{gaps} 
\end{center}
\end{figure}

When solving the eigenproblem of the pairing Hamiltonian $\hat{H}_{\rm
pair}$ we replace the standard BCS formalism by an approximate
projection of the BCS wave function on a given particle number
\cite{Go86}. This procedure is based on the gaussian overlap
approximation (GOA) of the generator coordinate method (GCM) and gives
results close to the exact projection. Moreover, the GCM projection is
very convenient because its results are easy to present in terms of the
usual BCS formalism. The form of BCS equations for the energy gap
$\Delta$ and the Fermi energy $\lambda$ remains unchanged. The only
effect of the particle number projection is that the standard BCS
energy (of particles with isospin projection~$t$) 
\bq\label{bcs}
V_{\rm BCS}(\Delta)= \sum_{\nu >0}(2e_{\nu}v_{\nu}^2-G_tv_{\nu}^4) 
-{\Delta^2\over G_t},
\eq
is now corrected so as that the pairing energy is 
\bq\label{pr}
V_{\rm pair}(\Delta)=V_{\rm BCS}(\Delta)-
\left( \sum_{\nu >0}\frac{\displaystyle (e_{\nu}-\lambda)(u_{\nu}^2-v_{\nu}^2)
+2\Delta u_{\nu}v_{\nu}+G_tv_{\nu}^4}{E_{\nu}^2}\right)
\left/ \left(\sum_{\nu >0}E_{\nu}^{-2}\right) \right.\; .
\eq
Here $e_{\nu}$ is an eigenvalue of $\hat{h}$ and $v_{\nu}^2,\, u_{\nu}^2 =
1-v_{\nu}^2$ are the usual BCS occupation probabilities, and
$E_{\nu}=\sqrt{(e_{\nu}-\lambda)^2+\Delta^2}$ is the quasiparticle 
energy. The $2\sqrt{15Z \mbox{ (or $N$)}}$ levels closest to the Fermi
surface were taken when solving the BCS equation. It was shown
\cite{Pi89} that when the projection on a given particle number is
performed the pairing strength parameter should be renormalized with
respect to the standard value in order to get the proper energy gaps.
With the estimation: 
\bq
G_t=\left\{\begin{array}{ll}
   g_{0}/(A{-}Z)^{2/3}&\ \ \mbox{ for }t=1/2,\\
   g_{0}/Z^{2/3} &\ \ \mbox{ for }t=-1/2
   \end{array}
   \right.
\eq
where $g_{0}=0.285\,\hbar\omega_{0}$, we can reasonably reproduce 
the experimental values of energy gaps. This is shown in Fig.~\ref{gaps} 
where experimental gaps deduced from measured or estimated \cite{Mo93}
nuclear masses are compared with theoretical values of neutron gap
$\Delta_n$. 

We apply
the Strutinsky microscopic-macroscopic method \cite{Ni69} to calculate
the collective potential and we use the cranking model for determining
moments of inertia, gyromagnetic functions and vibrational inertial
functions. The standard cranking formulae read \cite{Ka76,Ro77}: 
\bq\label{mom}
{\cal J}_k=2\hbar^2\sum_{\nu,\nu'>0}\frac{(u_{\nu}v_{\nu'} -u_{\nu'}v_{\nu})^2}
{E_{\nu}+E_{\nu'}}
|\langle\nu|\hat{j}_k|\nu'\rangle |^2 ,
\eq
\bq\label{gyr}
{\cal G}_k=\frac{2\hbar^2}{{\cal J}_k}\sum_{\nu,\nu'>0}\frac{(u_{\nu}v_{\nu'} 
-u_{\nu'}v_{\nu})^2}{E_{\nu}+E_{\nu'}}
\langle\nu|\hat{j}_k|\nu'\rangle \langle\nu'|\hat{\mu}_k|\nu\rangle \ ,
\eq
where $\hat{j}_k$ and $\hat{\mu}_k\; (k=1,\, 2,\, 3)$ are components of
the single particle total angular momentum and magnetic moment,
respectively, and 
\bqa\label{mas} 
\nonumber
&&B_{qq'}=\frac{\hbar^2}{f(q)f(q')}\left\{2\sum_{\nu,\nu'>0}
\frac{(u_{\nu}v_{\nu'}+u_{\nu'}v_{\nu})^2}{(E_{\nu}+E_{\nu'})^3}
 \langle\nu|\frac{\partial \hat{h}}{\partial q}|\nu'\rangle 
 \langle\nu'|\frac{\partial \hat{h}}{\partial q'}|\nu\rangle+\right.\\
&& + \left.\frac{1}{4}\sum_{\nu >0}\frac{\Delta^2}{E_{\nu}^5}
\left[\frac{\partial\lambda}{\partial q}\frac{\partial\lambda}{\partial q'}
-\left(\frac{\partial\lambda}{\partial q}\langle\nu|\frac{\partial \hat{h}}
{\partial q'}|\nu\rangle +
\frac{\partial\lambda}{\partial q'}\langle\nu|\frac{\partial \hat{h}}
{\partial q}|\nu\rangle \right)\right]\right\}
\eqa
for $q,\, q' = \B$ or $\G$, where $f(\B)=1$, $f(\G)=\B$.
To be complete we quote at the end formulae for the electric moments:
\bqa\label{el}
&Q_0=& Ze\sqrt{16\pi\over 5}\sum_{\nu>0}2v_{\nu}^2
\langle \nu |{1\over 2}(1-\tau_3)r^2Y_{20}|\nu\rangle \ ,\\
&Q_2=& Ze\sqrt{16\pi\over 5}\sum_{\nu>0}2v_{\nu}^2
\langle \nu |{1\over 2}(1-\tau_3){r^2\over \sqrt{2}}(Y_{22}+Y_{2-2})
|\nu\rangle \ ,\\
&R^2=&\sum_{\nu>0}2v_{\nu}^2\langle\nu|{1\over 2}(1-\tau_3)r^2|\nu\rangle \ .
\eqa

\section{Effect of the collective pairing vibrations}
\label{zppv}

It is suspected for a long time \cite{Ro77,St85,Pi93} that the pairing 
vibrations are coupled to the collective quadrupole motion. Here we do not
construct a complete "quadrupole + pairing" collective model which
would have nine degrees of freedom and could not be easy to solve. We
just take approximately into account the effect of the pairing
vibrations on the quadrupole excitations in question.

The collective pairing Hamiltonian has the following structure
\cite{Be70,Go85a,Do86}: 
\bqa\label{hampa}
\nonumber & \hat{\cal H}_{\rm pair}=&-\frac{\hbar^2}
{2\sqrt{w_{\Delta}(\Delta)r_{\Phi}(\Delta)}}\frac{\partial}
{\partial\Delta}\frac{\sqrt{w_{\Delta}(\Delta)r_{\Phi}(\Delta)}}
{B_{\Delta\Delta}(\Delta)}
\frac{\partial}{\partial\Delta} + \frac{\hbar^2}{2{\cal J}_{\Phi}(\Delta)}
\hat{\cal N}^2\\
&&+ A_{\rm pair}(\Delta)\hat{\cal N} +V_{\rm pair}(\Delta)
\eqa
where $\hat{\cal N}= -i{\partial}/{\partial \Phi}$ is the particle number 
excess operator and $\Phi$ is the gauge angle. Expressions for the
functions $B_{\Delta\Delta}(\Delta),\, {\cal J}_{\Phi}(\Delta),\,
w_{\Delta}(\Delta),\, r_{\Phi}(\Delta)$ and $A_{\rm pair}(\Delta)$
obtained from GCM and GOA are given in ref. \cite{Go85a}.

\begin{figure}[ht]
\begin{center}
\scalebox{0.5}{\includegraphics{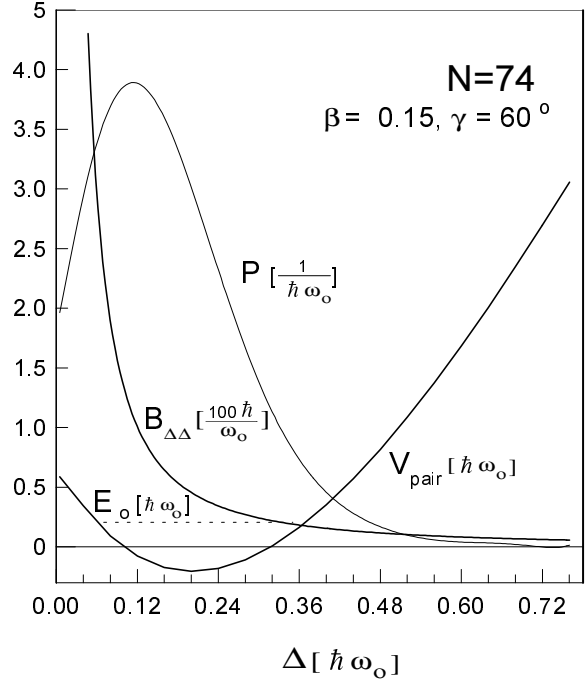}}
\caption{\footnotesize The neutron zero-point pairing vibration at deformation
$\B = 0.15,\, \G = 60^{\circ}$ for neutron number $N = 74$. The equilibrium
value of the energy gap $\Delta_{\rm BCS} \approx 0.20\hbar\omega_0$
while the most probable value $\Delta_0\approx
0.12\hbar\omega_0$.}\label{del0} 
\end{center}
\vspace{-2mm}
\end{figure}

\begin{figure}[hbt]
\begin{center}
\scalebox{0.4}{\includegraphics{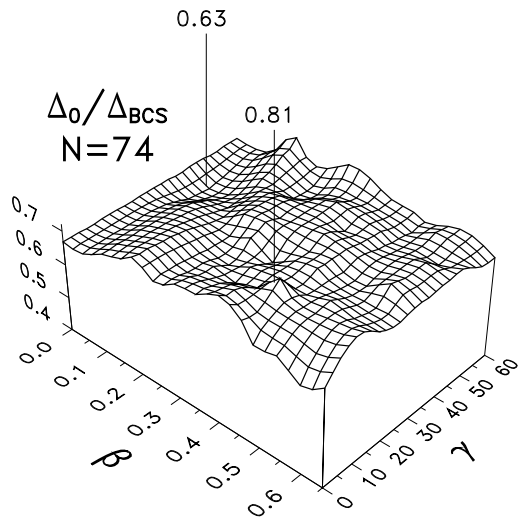}}
\caption{\footnotesize Ratio of the neutron energy gaps
$\Delta_0/\Delta_{\rm BCS}$ as a function of $\B$ and $\G$ for the
neutron number $N = 74$. The minimal and maximal values of the ratio
are marked on the surface.} \label{dbg}
\end{center}
\end{figure}

The ground-state energy, $E_0$, of the pairing vibrations is to be
found from the following eigenvalue problem:
\bq\label{zp}
\hat{\cal H}_{\rm pair}\Psi_0(\Delta)= E_0\Psi_0(\Delta),\quad 
\hat{\cal N}\Psi_0(\Delta) =0\ ,
\eq
which we solve at each deformation point $(\B,\, \G)$ for a given
number $N$ of neutrons or a given number $Z$ of protons, respectively.
This way we find $E_0(\B,\G)$ and $\Psi_0(\Delta;\B,\G)$ separately for
neutrons and protons in a given nucleus. Then, we replace $\Delta_{\rm
BCS}$ of the BCS (such that $V_{\rm pair}(\Delta)=\min$) in all
collective functions (eqs (\ref{mom}), (\ref{gyr}), (\ref{mas}) and
(\ref{el})) by the most probable value of the energy gap $\Delta_0$,
i.e. for which the probability 
\bq\label{prob}
P(\Delta)=|\Psi_0|^2w_{\Delta}r_{\Phi}
\eq
of finding a given value of $\Delta$ in the ground state is maximal 
(Fig.~\ref{del0}). Since $B_{\Delta\Delta}$ decreases rapidly with the 
increase of $\Delta$, the values of $\Delta_0$ are systematically lower
than the equilibrium energy gaps for all values of $\B$ and $\G$ (cf.
\cite{Pi93}). An example of such a behaviour is demonstrated in
Fig.~\ref{dbg}.

\section{Results}
\label{results}

The collective model in version presented in Section \ref{cm} with the 
Hamiltonian and electromagnetic multipole operators constructed in the
framework of microscopic model discussed in Section \ref{micr} has been
applied for a description of the collective states in even--even isotopes
of Te, Xe, Ba, Ce, Nd and Sm. The effect of coupling with the pairing
vibrations as reported in Section \ref{zppv} has been taken
approximately into account. The calculations have been performed with
no free parameters to be fitted. The results are discussed below.

\subsection{The energy spectra}
\label{levels}

\begin{figure}[htb]
\vspace*{6pt}
\begin{center}
\scalebox{0.55}{\includegraphics{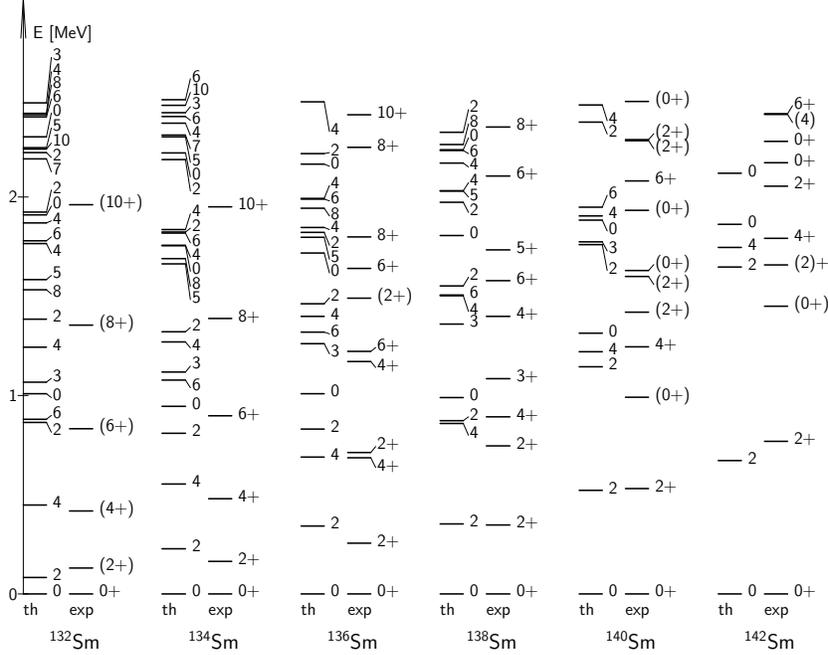}}
\vspace*{-5pt}
\caption{\footnotesize Calculated (th) and experimental (exp) energy
levels in the even--even samarium isotopes $^{132-142}$Sm. The
theoretical levels are marked with spin (parity of all of states is
positive), while the experimental ones --- with spin and parity.
Experimental data are taken from \cite{Taxx}.} \label{lsm}
\end{center}
\end{figure}

\begin{figure}[hbt]
\vspace*{6pt}
\begin{center}
\scalebox{0.55}{\includegraphics{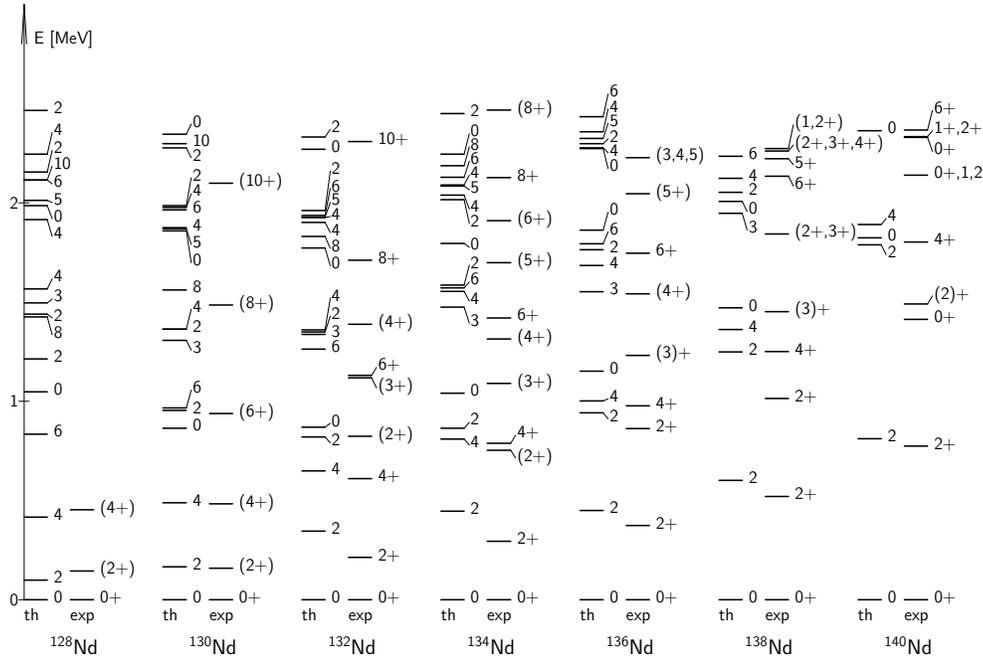}}
\vspace*{-5pt}
\caption{\footnotesize The same as in Fig.~\ref{lsm} but for neodymium
isotopes $^{126-140}$Nd.}\label{lnd}
\end{center}
\end{figure}

\begin{figure}[htb]
\vspace*{16pt}
\begin{center}
\scalebox{0.55}{\includegraphics{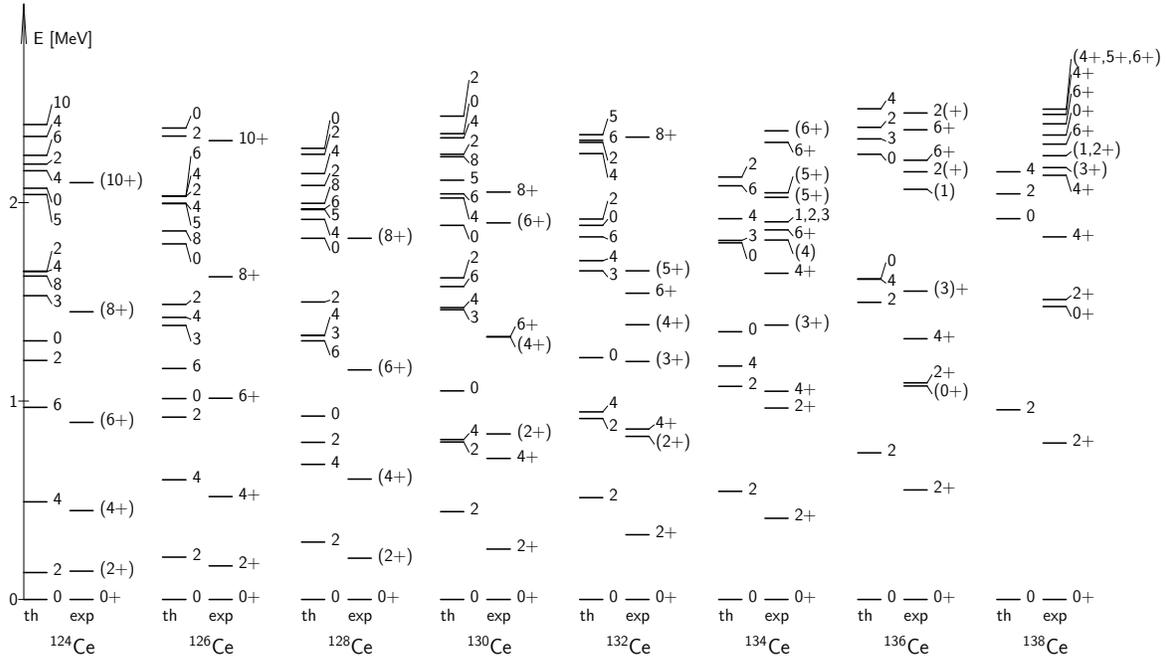}}
\caption{\footnotesize The same as in Fig.~\ref{lsm} but for cerium isotopes 
$^{124-138}$Ce.}\label{lce}
\end{center}
\end{figure}

\begin{figure}[htb]
\vspace*{16pt}
\begin{center}
\scalebox{0.55}{\includegraphics{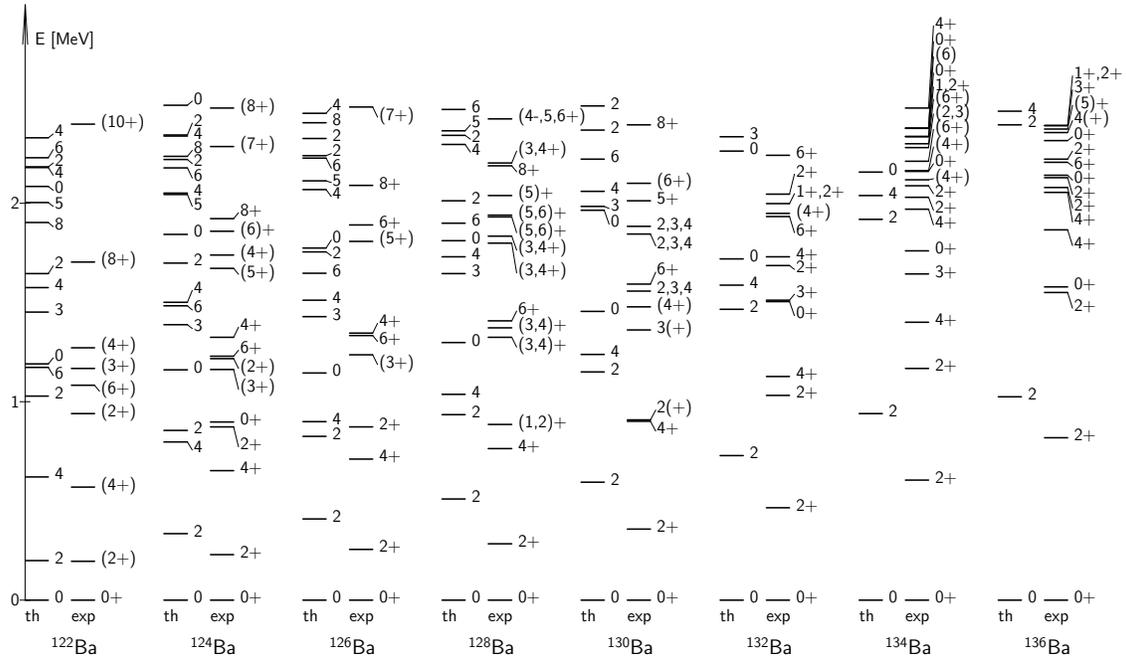}}
\caption{\footnotesize The same as in Fig.~\ref{lsm} but for barium isotopes 
$^{122-136}$Ba.}\label{lba}
\end{center}
\end{figure}

\begin{figure}[htb]
\vspace*{16pt}
\begin{center}
\scalebox{0.55}{\includegraphics{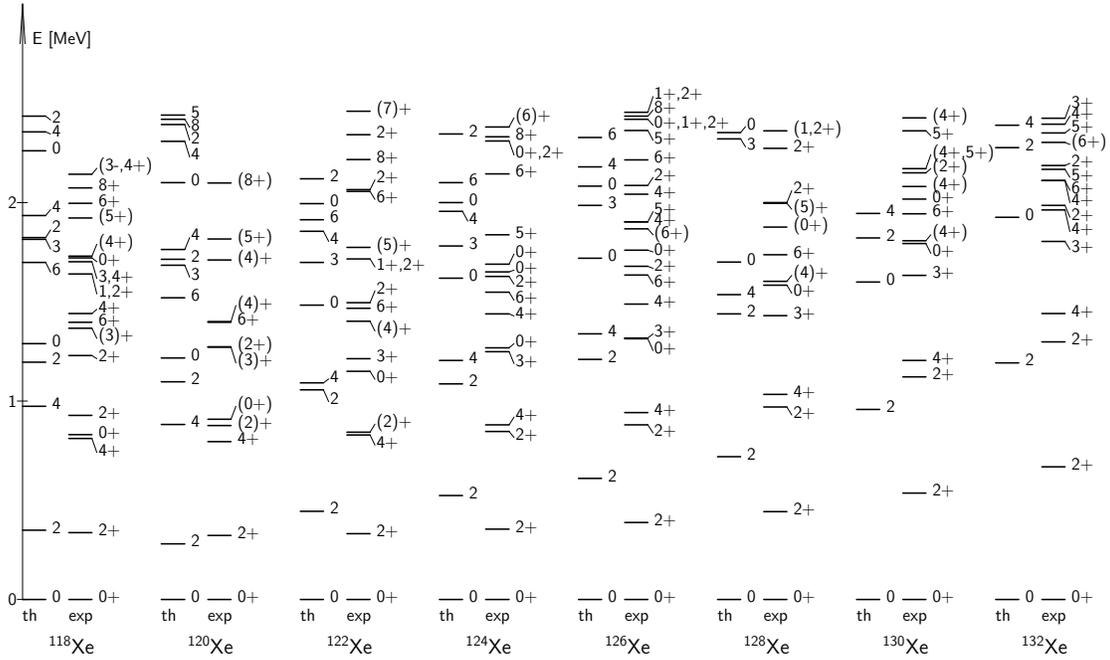}}
\caption{\footnotesize The same as in Fig.~\ref{lsm} but for xenon isotopes 
$^{118-132}$Xe.}\label{lxe}
\end{center}
\end{figure}

\begin{figure}[htb]
\vspace*{16pt}
\begin{center}
\scalebox{0.55}{\includegraphics{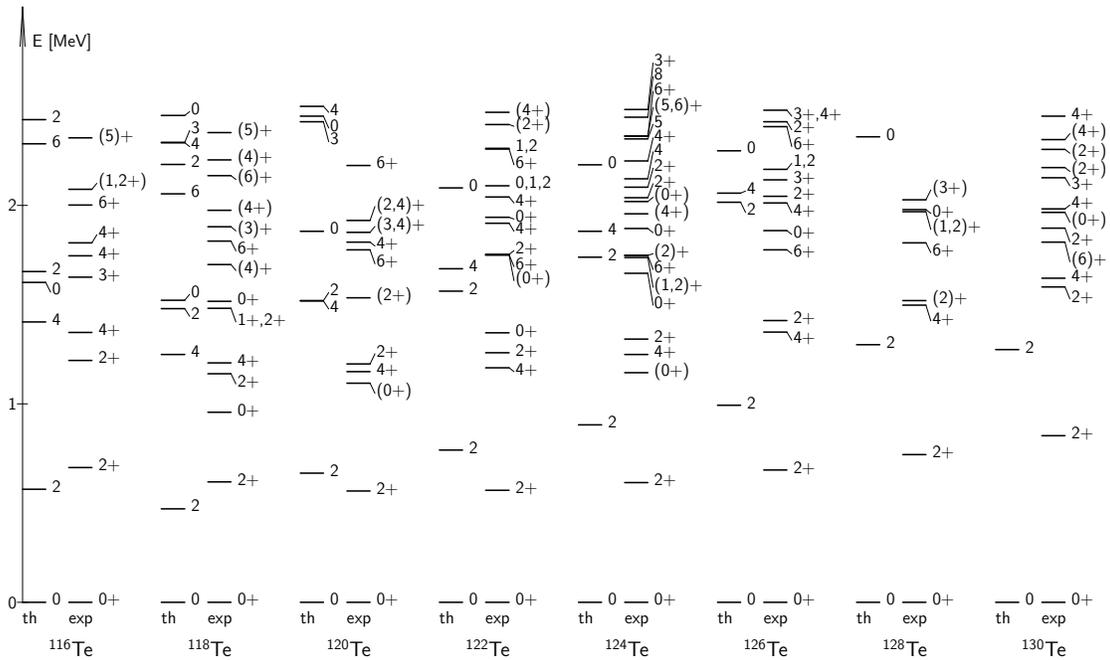}}
\caption{\footnotesize The same as in Fig.~\ref{lsm} but for tellurium
isotopes $^{116-130}$Te.}\label{lte}
\end{center}
\end{figure}

The calculated energy levels in the six isotope chains in question are
shown in Figs \ref{lsm}, \ref{lnd}, \ref{lce}, \ref{lba}, \ref{lxe} and 
\ref{lte}, respectively and compared with the experimental spectra
which are taken from \cite{Taxx} unless another reference is quoted.

\begin{figure}[htb]
\vspace*{16pt}
\begin{center}
\scalebox{0.55}{\includegraphics{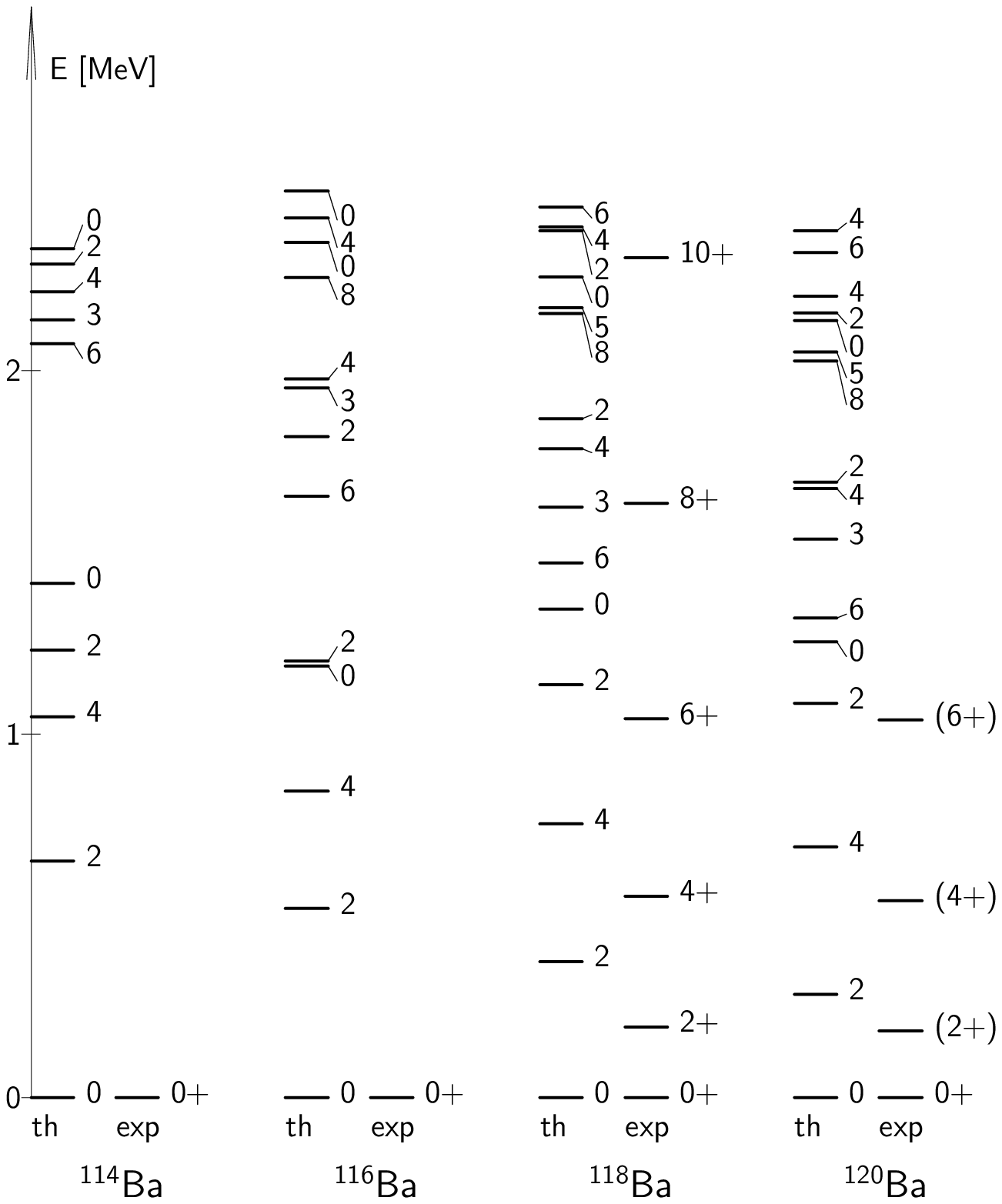}}
\caption{\footnotesize The same as in Fig.~\ref{lsm} but for light barium
isotopes $^{114-120}$Ba. Experimental data for $^{118}$Ba are taken
from \cite{newba}.}\label{lbali}
\end{center}
\end{figure}

The energy spectra in all of the isotopes of the two heaviest elements in
question, Nd and Sm, are reproduced fairly well. The yrast states $2^+,\, 
4^+,\, 6^+$ and $8^+$ and the second $2^+$ state are in more
or less correct positions, although the yrast bands are little bit too 
stretched (Figs \ref{lsm} and \ref{lnd}). Similar results we have got
for three lightest cerium isotopes, $^{124}$Ce, $^{126}$Ce and
$^{128}$Ce while the first $2^+$ state becomes too high for heavier 
isotopes. In the case of two heaviest ones, $^{136}$Ce and $^{138}$Ce, also 
two--phonon triplet of states $0^+,\, 2^+,\, 4^+$ is visibly too high (Fig.
\ref{lce}). The spectra only of two or three lightest isotopes of
barium, xenon and tellurium, $^{122}$Ba, $^{124}$Ba, $^{126}$Ba,
$^{118}$Xe, $^{120}$Xe, $^{116}$Te, $^{118}$Te, are of a good energy
scale. For the heavier isotopes the calculation gives too stretched
energy spectra (Figs \ref{lba}, \ref{lxe} and \ref{lte}). The
theoretical energy levels for very light barium isotopes $^{114-120}$Ba
are plotted additionally in Fig \ref{lbali}. The results for $^{118}$Ba
are compared with very recent experimental data from Ref. \cite{newba}.
Here the energy scale of the theoretical results for $^{118}$Ba and
$^{120}$Ba seems also to be too stretched with respect to the
experimental tendency. Probably the same occurs for unknown yet
levels in $^{114}$Ba and $^{116}$Ba which have recently been investigated 
for the first time \cite{Ja97}.

\subsection{The electromagnetic properties}
\label{elmg}

\begin{figure}[htb]
\begin{center}
\scalebox{0.5}{\includegraphics{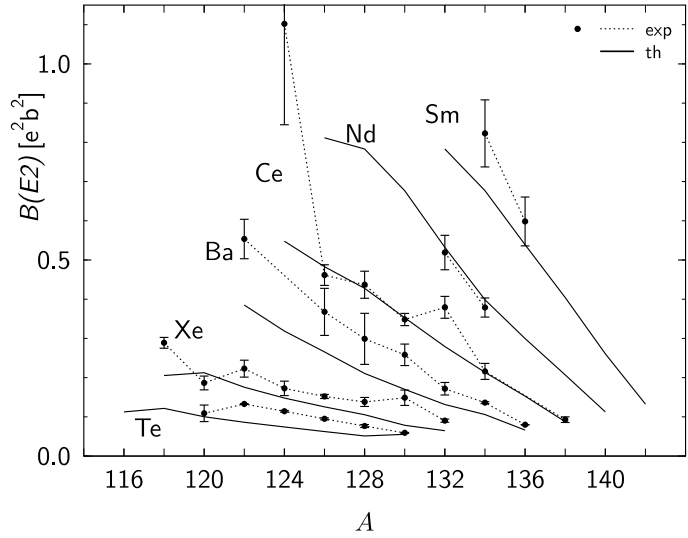}}
\caption{\footnotesize Calculated and experimental \cite{Taxx} reduced
transition probabilities \mbox{$B({\rm E2};2^+\rightarrow 0^+)$} (in
e$^2$b$^2$) from the first excited $2^+$ state to the ground
state.}\label{be2} 
\end{center}
\end{figure}

The results of calculation of the reduced transition probability from the 
first excited $2^+$ state to the ground state, $B({\rm E2}; 2^+\rightarrow
0^+)$ are compared with experimental data in Fig.~\ref{be2}. It is seen that
the $B$(E2) values for isotopes of Ce, Nd and Sm are reproduced quite well.
The calculated probabilities for most cases of the Xe and Ba isotopes are
only a little too large and a tendency to decrease probabilities with increase
the neutron number is reproduced correctly. An agreement of the theory with 
experiment is, in general, better for heavier Xe and Ba isotopes and worse for
lighter ones. A similar observation can be made for tellurium isotopes.
However, the theory gives $B$(E2) values plainly too big in this case
although the tendency of changes from isotope to isotope is again correct.

\begin{figure}[htb]
\begin{center}
\scalebox{0.5}{\includegraphics{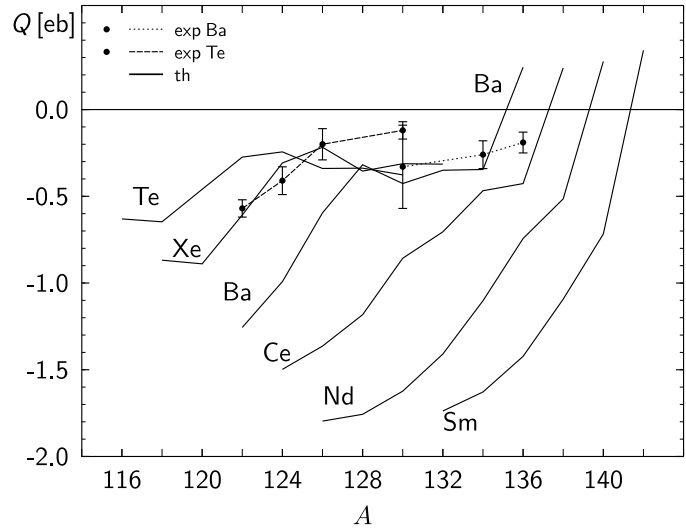}}
\caption{\footnotesize Calculated and experimental \cite{Taxx}
spectroscopic electric quadrupole moments (in eb) of the first excited
$2^+$ state.}\label{qm} 
\end{center}
\end{figure}

Theoretical values of the spectroscopic quadrupole moment of the first $2^+$
state are shown in Fig.~\ref{qm}. The experimental data are scanty for this
quantity and it is difficult to estimate of agreement between calculations
and measurements. In cases of $^{130}$Ba and $^{134}$Ba it is very good. An
opposite sign of the quadrupole moment is predicted for $^{136}$Ba. For the
Te isotopes the agreement is rather poor.

\begin{figure}[htb]
\begin{center}
\scalebox{0.5}{\includegraphics{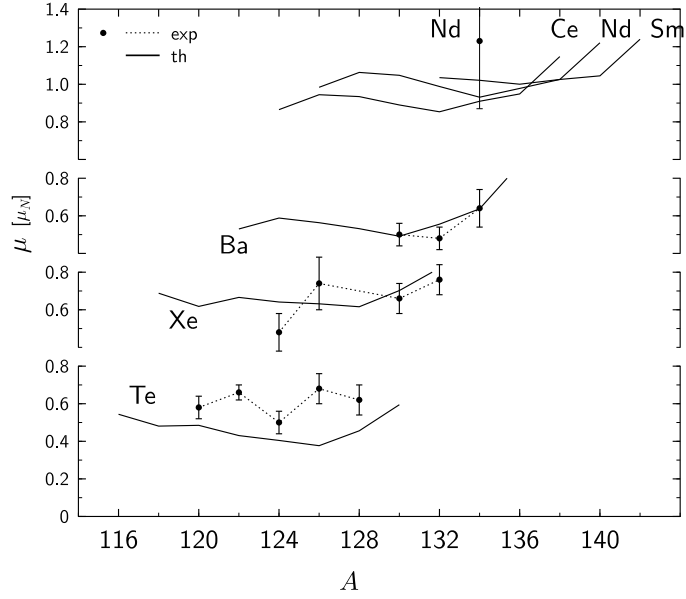}}
\caption{\footnotesize Calculated and experimental \cite{Taxx} magnetic dipole
moments (in $\mu_{\rm N}$) of the first excited $2^+$ state.}\label{mu}
\end{center}
\end{figure}

The calculated magnetic dipole moments agree with the experimental values
in all of known cases of the Xe, Ba and Nd isotopes. The agreement between
the theory and experiment for the tellurium isotopes is again poorer.
This is shown in Fig.~\ref{mu}.

\begin{figure}[htb]
\vspace{-2mm}
\begin{center}
\scalebox{0.5}{\includegraphics{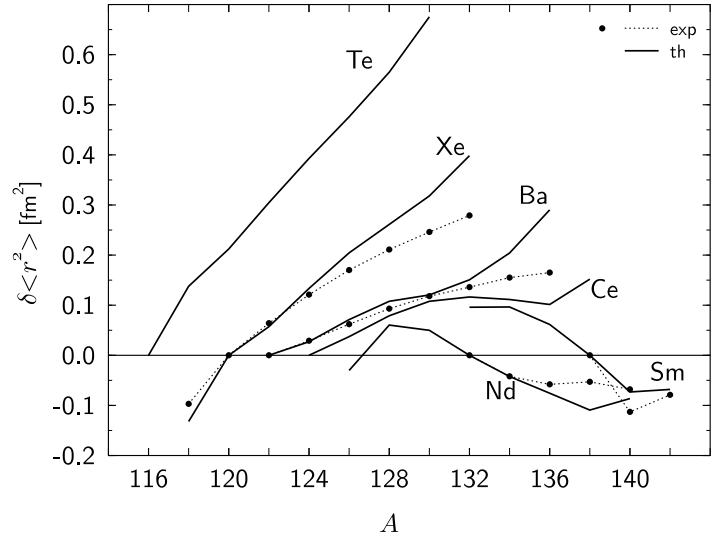}}
\caption{\footnotesize Calculated and experimental \cite{Otten} isotopic shifts
of the mean square radii in the ground states.}\label{r2}
\end{center}
\end{figure}

Fig.~\ref{r2} shows the results of calculation of isotopic shifts for the
charge root mean square radii. The experimental values in all of cases
available for the Xe and Ba isotopes \cite{Otten} are reproduced very well. 

The knowledge of experimental data for other electromagnetic properties like 
transition probabilities from higher excited states, branching and
mixing ratios is still rather poor in the investigated region of
nuclear chart and, therefore, we do not discuss them here. 

\section{Conclusions}
\label{concl}

The low-lying quadrupole collective states of 48 even-even nuclei from
the region $50<Z,\, N<82$ of the chart of nuclides are investigated within
the framework of the generalized Bohr Hamiltonian. The collective potential
and inertial functions are determined microscopically. There are no
free parameters in the calculation. Taking the coupling of the
quadrupole and pairing vibrations into account brings the energy levels
down to the scale comparable with that of experimental one. In general,
the GBH works better in the case of elements with larger $Z$, namely,
Ce, Nd and Sm than for Xe, Ba and, especially, Te which has only two
protons outside the closed shell $Z=50$. Energies are better reproduced
by the calculation for isotopes with lower number of neutrons. For those
with neutron number $N=78,\, 80$ the energy levels are, as a rule, too high.
On the contrary, electromagnetic properties seem to be better reproduced just
in the case of heavier isotopes. The calculated reduced probability of the 
$2^+\rightarrow 0^+$ transition, if not reproduced well, has a tendency to be
too large, especially in cases of lighter isotopes for which energy levels
lie in more or less correct position.

It does not make sense to discuss how good or how bad are our results 
obtained with no free parameters in comparison to those of other
approaches in which some parameters are usually fitted to the data. 

Finally, we may say that though the collective model based on the generalized 
Bohr Hamiltonian gives qualitatively reasonable results for nuclei from
the major shell $50< Z,\, N<82$, it is not able to reproduce all of collective 
characteristics simultaneously in detail. We think that dynamical effects of 
coupling with pairing vibrations can improve the description of
collective states in question. 
 
\section*{Appendix.\\
 Symmetrization under the O group}

All of transformations which change names and senses of the intrinsic axes
and form the octahedral group O can be expressed through the three well
known Bohr rotations, $R_1,\, R_2,\, R_3$ (see refs 
\cite{Bo52,Ku67,EG87,Co76} for their definition). Therefore, it is
sufficient to symmetrize functions $\varphi^{IM}_{Lm}(\B,\G,\Omega)$
of eq. (\ref{set1}) with respect to $R_i,\, (i=1,\, 2,\, 3)$. Symmetrization
under $R_1$ and $R_2$ gives \bq\label{set2}
(1+R_1)(1+R_2+R^2_2+R^3_2){\varphi}^{IM}_{Lmn}(\B,\G,\Omega)\propto
\psi^{IM}_{Lmn}(\B,\G,\Omega)=e^{-
\mu\B^2/2}\B^{n}p^{}_{Lm}(\G)\Phi^{I}_{ML}(\Omega)
\eq
where
\bq\label{pe1}
p^{}_{Lm}(\G)=\left\{ \begin{array}{l}
\cos m\G \quad \mbox{for}\ \ L/2 \ \ \mbox{even},\\
\sin m\G \quad \mbox{for}\ \ L/2 \ \ \mbox{odd}
\end{array} \right.
\eq
and $\Phi^I_{ML}$ is given by eq. (\ref{symD}). 
The additional symmetrization under $R_3$ leads to the functions:
\bq\label{set3}
(1+R_3+R^2_3)\psi^{IM}_{Lmn}(\B,\G,\Omega)=
\Psi^{IM}_{Lmn}(\B,\G,\Omega)=
e^{-\mu\B^2/2}\B^{n}\sum_{K={\rm even}\ge 0}
f^{I}_{LmK}(\G)\Phi^{I}_{MK}(\Omega)
\eq
where
\bq\label{f}
f^{I}_{LmK}(\G)=c^{I}_{LmK}p^{}_{Km}(\G).
\eq
Coefficients
$c^I_{LmK}$ are equal to
\bqa
 \nn c^{I}_{LmK}=\delta^{}_{LK}+2(-
1)^{K/2}[(1+\delta^{}_{L0})(1+\delta^{}_{K0})]^{-1/2}
d^{I}_{KL}(\pi/2)a^{}_{LmK}, &&\\[4mm]
a_{LmK}=\left\{ \begin{array}{rl}
2\cos(2m\pi/3) &\quad \mbox{for }\; (K+L)/2 \ \mbox{even},\\
2\sin(2m\pi/3) &\quad \mbox{for }\; K/2 \ \mbox{even, } L/2 \ \mbox{odd},\\
-2\sin(2m\pi/3) &\quad \mbox{for }\; K/2 \ \mbox{odd, } L/2 \ \mbox{even}.
\end{array} 
\right. &&
\eqa
Function $d^I_{KL}(\theta)$ is related to the Wigner function in a usual way:
$D^{I}_{KL}(\phi,\theta,\psi)=
e^{iK\phi}d^{I}_{KL}(\theta)e^{iL\psi}$. 

Since $R_3$ does not commute with $R_1$ and $R_2$ the set of functions of
eq. (\ref{set3}) is linear dependent. Selection of linear independent
functions from the set (\ref{set3}) leads to restrictions on values of
quantum the number $L$ for given $I$ and $m$. The number $L$ takes even
integer values ranging from $2\delta_{I\, {\rm odd}}=2(1 - \delta_{I\,
{\rm even}})$ to $L^I_{\rm max}(m)$, where $\delta_{I\, {\rm even}}$
is equal to 1 for an even $I$ and 0 for an odd one, and 
\bq\label{lmax}
L^I_{\rm max}(m)=\left\{ \begin{array}{lcl}
2\cdot [I/12] + 2\Delta_I - 2\delIe & \mbox{~~for~~}&m = 0,\\
2\cdot [I/6]      & \mbox{~~for~~}&m>0,\, I\mod 3 = 1,\\
2\cdot [I/6] - 2\delIe & \mbox{~~for~~}& m>0,\, m\mod 3 =0,\, I\mod 3 = 2,\\
                & & {\rm or}\; m\mod 3 \neq 0, I\mod 3 = 0,\\
2\cdot[I/6] + 2\delta_{I\, {\rm odd}} &\mbox{~~for~~}& m>0,\, m\mod 3 = 0,\, 
I\mod 3 = 0,\\
                & & {\rm or}\; m\mod 3 \neq 0,\, I\mod 3 = 2,\\
\end{array}
\right.
\eq
with
\bq
\Delta_I = \left\{ \begin{array}{ll}
1& \mbox{~~for~~} I\mod 12 = 0,\, 4,\, 6,\, 8,\, 9,\, 10,\\
0 &\mbox{~~for~~} I\mod 12 = 1,\, 2,\, 3,\, 5,\, 7,\, 11.\\
\end{array}
\right.
\eq
A rational number $N/M$ in square brackets, $[N/M]$, stands for its
integer part. 

We have to consider also the boundary conditions in $\G$ (it is for
$\G=k \pi/3$). The combinations of the linear independent functions
$\Psi^{IM}_{Lmn}$ which fulfill these conditions and, thus, form a
(nonorthogonal) basis in the domain of the Bohr 
Hamiltonian, are (cf. \cite{Pr98}):
\bq
\label{set4a}
\tilde{\Psi}^{IM}_{Lmn}(\B,\G,\Omega)=
e^{-\mu\B^2/2}\B^{n}\sum_{K={\rm even}\ge 0}
\tilde{f}^{I}_{LmK}(\G)\Phi^{I}_{MK}
\eq
where $\tilde{f}^I_{LmK}$ are the following combinations of functions
$f^I_{LmK}$ of eq. (\ref{f}) for all possible values of $K$:
\bq \label{kombf}
\begin{array}{lcl}
\multicolumn{3}{c}{\underline{m < 6}}\\
\tilde{f}^I_{00K} = f^I_{00K} & &\\
\tilde{f}^I_{02K} = f^I_{02K}+{1\over 2}f^{I}_{00K}& & \\
\tilde{f}^{I}_{L3K}=f^{I}_{L3K}+2\delta^{}_{L0}f^{I}_{01K}& \mbox{ for }&
L=(0),2,6,10,\ldots,4\cdot [(L^I_{\rm max}(3)+2)/4]-2\\
\tilde{f}^{I}_{L4K}=f^{I}_{L4K}-(-1)^{L/2}f^{I}_{L2K}& \mbox{ for } & 
L=(0),2,4,\ldots,L^I_{\rm max}(2)\\
\tilde{f}^{I}_{L5K}=f^{I}_{L5K}-(-1)^{L/2}f^{I}_{L1K}& \mbox{ for }&
 L=(0),2,4,\dots,L^I_{\rm max}(1)\\
 & & \\
\multicolumn{3}{c}{\underline{m\ge 6}}\\
\ \ m \mod 3 \ne 0& &\\
\tilde{f}^{I}_{LmK}=f^{I}_{LmK}-f^{I}_{Lm-6K}& \mbox{ for }& 
 L=(0),2,4,\ldots,L^I_{\rm max}(m)\\
\ \ m \mod 3 = 0& &\\
\tilde{f}^{I}_{LmK}=f^{I}_{LmK}-\delta_{L\mod 4\, 0}f^{I}_{Lm-6K}&
\mbox{ for }& L=(0),2,4,\ldots,L^I_{\rm max}(m)\\
\end{array}
\eq
where the lowest value of $L$ in parentheses is omitted for odd $I$'s.

The orthonormal (with respect to the measure ${\rm d}\tau_0$) set of
functions reads: 
\bq\label{orto}
\tilde{F}^{IM}_{Lmn}(\B,\G,\Omega)=e^{-\mu\B^2/2}R^{}_{nm}(\B ; \mu)
\tilde{T}^{IM}_{Lm}(\G,\Omega)
\eq
where
\bq
\label{orto1}
\tilde{T}^{IM}_{Lm}(\G,\Omega)=\sum_{K}\sum_{L'm'}E^{(Ip)}_{Lm,L'm'}
{\tilde{f}}^{I}_{L'm'K}(\G)\Phi^I_{MK}(\Omega)
\eq
and
\bq\label{orto2} 
R^{}_{nm}(\B ;\mu)=\sum_{n'}S^{(m)}_{nn'}(\mu)\B^{n'}
\eq
Lower diagonal matrices $S^{(m)}(\mu)$ and $E^{(p)}$ are obtained
through Cholesky-Bana\-chie\-wicz method. The running index $m'$ in eq.
(\ref{orto1}) has the same parity $p$ as that of $m$. So, we have two
matrices $E^{(Ip)}$, one for each parity, for a given $I$. On the other
hand, matrices $S^{(m)}(\mu)$ depend on $m$ itself. They depend also on
parameter $\mu$. However, it turns out that 
\bq
R_{nm}(\B;\mu)=\mu^{5/2}R^{}_{nm}(\mu\B;\mu{=}1)
\eq
Hence, it is sufficient to perform the orthonormalization only once.

Hamiltonian $\hat{\cal H}_{\rm coll}$ is hermitian with the measure 
${\rm d}\tau$ of eq. (\ref{meso}). The basis orthonormal with respect
to this measure is finally of the following form:
\bq\label{set5}
F^{IM}_{Lmn}(\B,\G,\Omega)
=(w(\B,\G)r(\B,\G))^{-1/4}\tilde{F}^{IM}_{Lmn}(\B,\G,\Omega).
\eq

\section*{Acknowledgement}
This work was supported in part by the Polish Committee for Scientific 
Research under Contract No. 2~P03B~068~13. 

\end{document}